\documentclass[runningheads]{llncs}
\usepackage{graphicx}
\usepackage{xcolor}
\usepackage[colorlinks,linkcolor=blue,citecolor=blue]{hyperref}

\usepackage{multirow,booktabs,amsmath,url,float}
\usepackage[T2A,T1]{fontenc}
\usepackage[utf8]{inputenc}
\usepackage[russian,english]{babel}

\begin{document}
\title{A Study of Multilingual End-to-End Speech Recognition for Kazakh, Russian, and English}
%
\titlerunning{A Study of Multilingual End-to-End Speech Recognition}
%
\author{Saida Mussakhojayeva \and Yerbolat Khassanov \and Huseyin Atakan Varol}
\authorrunning{S. Mussakhojayeva et al.}
%
\institute{Institute of Smart Systems and Artificial Intelligence (ISSAI),\\ Nazarbayev University, Nur-Sultan, Kazakhstan\\
\email{\{saida.mussakhojayeva,yerbolat.khassanov,ahvarol\}@nu.edu.kz}}
\maketitle              

\begin{abstract}
We study training a single end-to-end (E2E) automatic speech recognition (ASR) model for three languages used in Kazakhstan: Kazakh, Russian, and English.
We first describe the development of multilingual E2E ASR based on Transformer networks and then perform an extensive assessment on the aforementioned languages.
We also compare two variants of output grapheme set construction: combined and independent.
Furthermore, we evaluate the impact of LMs and data augmentation techniques on the recognition performance of the multilingual E2E ASR.
In addition, we present several datasets for training and evaluation purposes.
Experiment results show that the multilingual models achieve comparable performances to the monolingual baselines with a similar number of parameters.
Our best monolingual and multilingual models achieved 20.9\% and 20.5\% average word error rates on the combined test set, respectively.
To ensure the reproducibility of our experiments and results, we share our training recipes, datasets, and pre-trained models\footnote{\label{ft:github}\url{https://github.com/IS2AI/MultilingualASR}}.

\keywords{Speech recognition \and Multilingual \and End-to-end \and Dataset \and Transformer \and Kazakh \and Russian \and English.}
\end{abstract}
%
%
%
\section{Introduction}
This work aims to study the effectiveness of a multilingual end-to-end (E2E) automatic speech recognition (ASR) system applied to three languages used in Kazakhstan: Kazakh, Russian, and English.
Kazakhstan is a multinational country where Kazakh is the official state language, whereas Russian and English are the languages of interethnic and international communication commonly used in business, science, and education. 
These three languages are part of a large-scale cultural project initiated by the government named ``The Trinity of Languages''~\cite{Trinity}.
The goal of the project is the mastery of the aforementioned languages by the Kazakhstani people.
This will presumably enable citizens' successful integration into the international economic and scientific environments.
In this regard, we initiate the first study of a single joint E2E ASR model applied to simultaneously recognize the Kazakh, Russian, and English languages.

Having a single ASR model for multiple languages considerably simplifies training, deployment, and maintenance~\cite{DBLP:conf/interspeech/PratapSTHLSC20}.
In particular, this is advantageous for multilingual communities where several languages are used for communication.
A multilingual ASR system can automatically detect an input language and produce corresponding transcripts without prompting for language or requiring visual and tactile interfaces.
This becomes especially useful when ASR is employed in a pipeline of a larger system, such as message dictation, voice command recognition, virtual assistants, a transcription engine on online audio/video sharing platforms (e.g., YouTube), and so on.

Recently presented E2E ASR architectures have been shown to be effective for the multilingual speech recognition task~\cite{DBLP:conf/slt/ChoBLWMYKWH18,DBLP:journals/corr/abs-2104-14830,DBLP:conf/icassp/ToshniwalSWLMWR18}.
It has also been demonstrated that the E2E approaches achieve comparable results to the conventional deep neural network-hidden Markov model (DNN-HMM) ASR~\cite{DBLP:conf/icml/GravesJ14,DBLP:journals/corr/HannunCCCDEPSSCN14}.
Moreover, they significantly reduce the burden of developing ASR systems thanks to the encapsulation of the acoustic, pronunciation, and language models under a single network.
Importantly, the E2E ASR models obviate the need for a pronunciation model, which requires significant human effort to be constructed and often proves critical to overall performance~\cite{DBLP:conf/icml/GravesJ14}. 
All these features make the E2E architectures attractive for the multilingual speech recognition task.

This work leverages the above-mentioned advantages and studies the multilingual E2E ASR systems applied to simultaneously recognize the Kazakh, Russian, and English languages.
Specifically, we thoroughly explore the performance of the Transformer-based E2E architecture~\cite{DBLP:conf/nips/VaswaniSPUJGKP17}.
To the best of our knowledge, this is the first study of multilingual E2E ASR dedicated to these languages.
We also compared the use of two different grapheme set construction methods (i.e., combined and independent).
We also analyzed the impact of language models and data augmentation techniques, such as speed perturbation~\cite{DBLP:conf/interspeech/KoPPK15} and spectral augmentation~\cite{DBLP:conf/interspeech/ParkCZCZCL19}.
We found that the multilingual models can achieve comparable results to strong monolingual baselines, despite having a similar number of parameters.
To enable experiment reproducibility and facilitate future research, we share our training recipes, datasets, and pre-trained models\textsuperscript{\ref{ft:github}}.

Besides conducting the first detailed study of multilingual E2E ASR for Kazakh, Russian, and English, other contributions of this paper are:
\begin{itemize}
    \item We introduce a 7-hour evaluation set of transcribed Kazakh-accented English audio recordings (i.e., native Kazakh speakers reading English sentences extracted from the SpeakingFaces dataset~\cite{DBLP:journals/corr/abs-2012-02961}).
    \item We introduce a 334-hour manually-cleaned subset of the OpenSTT dataset~\cite{OpenSTT} for the Russian language, which can also be used to train robust standalone Russian ASR systems.
\end{itemize}

The rest of the paper is organized as follows:
Section~\ref{sec:related} briefly reviews related works on multilingual ASR.
Sections~\ref{sec:models} and~\ref{sec:data} describe the multilingual models and datasets used in our experiments, respectively.
Section~\ref{sec:exp} presents the experimental setup and obtained results.
Section~\ref{sec:discus} discusses the important findings and highlights potential future work.
Finally, Section~\ref{sec:conclude} concludes this paper.

\section{Related Works}\label{sec:related}

A single model capable of recognizing multiple languages has been a long-term goal of the speech recognition community and remains an active area of research for decades~\cite{DBLP:conf/slt/ChoBLWMYKWH18,DBLP:conf/interspeech/PratapSTHLSC20,DBLP:conf/icassp/ToshniwalSWLMWR18}.
The use of a single model for several languages simplifies the ASR production pipeline significantly, since maintaining one model per language becomes cumbersome as the number of languages increases.
Furthermore, multilingual ASR systems leverage cross-lingual knowledge transfer, which has been shown to improve recognition performance, especially for low-resource languages~\cite{DBLP:journals/speech/BesacierBKS14a}.

Prior works on multilingual ASR have explored both hybrid DNN-HMM~\cite{DBLP:conf/icassp/GhoshalSR13} and E2E~\cite{DBLP:conf/icassp/ToshniwalSWLMWR18} architectures.
Both small- and large-capacity multilingual models with up to ten billion parameters have been studied~\cite{DBLP:journals/corr/abs-2104-14830}.
Offline and streaming speech recognition modes of multilingual ASR have also been investigated~\cite{DBLP:conf/interspeech/KannanDSWRWBCL19}. 
The authors of~\cite{DBLP:conf/naacl/AdamsWWY19} developed multilingual models capable of recognizing over 100 languages simultaneously. 
The prior works have also studied different approaches to further improve the multilingual speech recognition performance, such as multi-task~\cite{DBLP:conf/interspeech/HouDZYSS20} and transfer~\cite{DBLP:conf/slt/ChoBLWMYKWH18} learning.
In multi-task learning, a model is jointly trained with other tasks, such as language identification (LID), whereas, in transfer learning, a model pre-trained on other languages (usually high-resource ones) is fully or partially fine-tuned using the target languages.
However, to the best of our knowledge, there is no prior work dedicated to simultaneous recognition of the Kazakh, Russian, and English languages.

Among the aforementioned three languages, Russian and English are considered resource-rich, i.e., a large number of annotated datasets exist~\cite{VoxForge,DBLP:conf/lrec/ArdilaBDKMHMSTW20,mailabs} and extensive studies have been conducted, both in monolingual and multilingual settings~\cite{DBLP:conf/naacl/AdamsWWY19,markovnikov2017deep,DBLP:conf/interspeech/PratapSTHLSC20}.
On the other hand, Kazakh is considered a low-resource language, where annotated datasets and speech processing research have emerged only in recent years~\cite{khassanov-etal-2021-crowdsourced,DBLP:journals/corr/abs-2104-08459}.
The authors of~\cite{khassanov-etal-2021-crowdsourced} presented the first crowdsourced open-source Kazakh speech corpus and conducted initial Kazakh speech recognition experiments on both DNN-HMM and E2E architectures.
Similarly, the authors of~\cite{DBLP:journals/corr/abs-2104-08459} presented the first publicly available speech synthesis dataset for Kazakh.
Previously, the Kazakh language was part of several multilingual studies under the IARPA's Babel project~\cite{DBLP:conf/icassp/DalmiaSMB18,DBLP:conf/interspeech/HouDZYSS20,DBLP:conf/interspeech/KarafiatBMVGBC17}, and it was also explored in the context of Kazakh-Russian~\cite{DBLP:conf/specom/KhomitsevichMTR15,Ubskii2020ImpactOU} and Kazakh-English~\cite{akynova2014english} code-switching.

This work is the first to study the multilingual E2E ASR systems dedicated to Kazakh, Russian, and English, which we believe will further progress the speech processing research and advance the speech-enabled technology in Kazakhstan and its neighboring countries. These languages belong to different language families (i.e., Kazakh belongs to Turkic, Russian to Slavic, and English to Germanic), which poses an additional challenge to our work.
Therefore, we posit that our work will be of interest to the general speech research community, especially for researchers from the post-Soviet states, where Russian and English are also commonly spoken as lingua francas.

\section{Speech Recognition Models}\label{sec:models}

In our experiments, we consider three languages $(\mathcal{L}_{kz},\mathcal{L}_{ru},\mathcal{L}_{en})$, each with a corresponding grapheme set $(\mathcal{G}_{kz},\mathcal{G}_{ru},\mathcal{G}_{en})$\footnote{Note that the Kazakh and Russian grapheme sets overlap since they both use Cyrillic script.}.
In addition, each language has its own independent training set $(\{\mathcal{X}_{kz},\mathcal{Y}_{kz}\},\{\mathcal{X}_{ru},\mathcal{Y}_{ru}\},\{\mathcal{X}_{en},\mathcal{Y}_{en})\})$, where $\mathcal{X}$ is an input sequence of acoustic features and $\mathcal{Y}$ is a corresponding target sequence.

The training dataset for the multilingual models is constructed by combining all the three datasets without any form of re-weighting or re-balancing:
\begin{equation}
    \{\mathcal{X}_{all},\mathcal{Y}_{all}\}=\{\mathcal{X}_{kz},\mathcal{Y}_{kz}\}\cup\{\mathcal{X}_{ru},\mathcal{Y}_{ru})\cup\{\mathcal{X}_{en},\mathcal{Y}_{en}\}
\end{equation}
and the grapheme set for the combined dataset is similarly obtained as follows:
\begin{equation}
    \mathcal{G}_{all}=\mathcal{G}_{kz}\cup\mathcal{G}_{ru}\cup\mathcal{G}_{en}
\end{equation}

\subsection{Monolingual Model}
We begin by training randomly-initialized monolingual ASR models for each language using the corresponding training data $\{\mathcal{X}_{i},\mathcal{Y}_{i}\}$ and grapheme set $\mathcal{G}_{i}$, where $i\in(kz,ru,en)$.
These models are encoder-decoder networks based on the Transformer architecture~\cite{DBLP:conf/nips/VaswaniSPUJGKP17} and they will be used as a baseline.

\subsection{Multilingual Model}
Next, we train a joint model using the multilingual dataset $\{\mathcal{X}_{all},\mathcal{Y}_{all}\}$ and combined grapheme set $\mathcal{G}_{all}$.
The joint model is also based on the Transformer architecture; however, it is a single model whose parameters are shared across all three languages.
This model is not given any explicit indication that the training dataset is composed of multiple languages.
For the sake of fair comparison, we set the number of parameters and the structure of multilingual and monolingual models to be similar.

\subsubsection{Independent Grapheme Set.}
To alleviate the impact of language confusion in multilingual models, we explored the joint model trained using the independent grapheme sets.
To achieve this, we appended each character with the corresponding language code as follows:
\begin{itemize}
    \item[]Kazakh: `\textit{\foreignlanguage{russian}{с ә л е м}}' $\rightarrow$ `\textit{\foreignlanguage{russian}{с\_kz ә\_kz л\_kz е\_kz м\_kz}}'
    \item[]Russian: `\textit{\foreignlanguage{russian}{п р и в е т}}' $\rightarrow$ `\textit{\foreignlanguage{russian}{п\_ru р\_ru и\_ru в\_ru е\_ru т\_ru}}'
    \item[]English: `\textit{h e l l o}' $\rightarrow$ `\textit{h\_en e\_en l\_en l\_en o\_en}'
\end{itemize}

The training procedure and structure of the independent grapheme set joint model is similar to the standard joint model, except the output layer size is increased from $|\mathcal{G}_{all}|$ to $|\mathcal{G}_{kz}|+|\mathcal{G}_{ru}|+|\mathcal{G}_{en}|$. 

\section{Datasets}\label{sec:data}
To conduct multilingual speech recognition experiments, we used three datasets corresponding to the Kazakh, Russian, and English languages.
The dataset specifications are provided in Table~\ref{tab:dataset}.
To diminish the performance degradation caused by the challenges peculiar to unbalanced data~\cite{DBLP:journals/corr/abs-2104-14830}, we made the training set sizes of languages similar (in terms of duration).
Additionally, all audio recordings were resampled to 16 kHz and 16-bit format prior to training and evaluation. 
All datasets used in our experiments are available in our GitHub repository\textsuperscript{\ref{ft:github}}.

\begin{table}[t]
    \caption{The dataset statistics for the Kazakh, Russian, and English languages. Utterance and word counts are in thousands (k) or millions (M), and durations are in hours (hr). The overall statistics `Total' are obtained by combining the training, validation, and test sets across all the languages.}\label{tab:dataset}
    \renewcommand\arraystretch{1.2}
    \setlength{\tabcolsep}{1.5mm}
    \begin{tabular}{l|l|c|ccc}
        \toprule
        \multicolumn{2}{l|}{\textbf{Languages}}         & \textbf{Corpora}              & \textbf{Duration}     & \textbf{Utterances}   & \textbf{Words}\\
        \midrule
        \multirow{3}{*}{Kazakh}         & train         & \multirow{3}{*}{KSC~\cite{khassanov-etal-2021-crowdsourced}}  & 318.4 hr  & 147.2k    & 1.6M \\
                                        & valid         &                               & 7.1 hr                & 3.3k                 & 35.3k \\
                                        & test          &                               & 7.1 hr                & 3.3k                 & 35.9k \\\hline
        \multirow{5}{*}{Russian}        & train         & \multirow{2}{*}{OpenSTT-CS334}& 327.1 hr              & 223.0k               & 2.3M \\
                                        & valid         &                               & 7.1 hr                & 4.8k                 & 48.3k \\\cline{3-3}
                                        & test-B (books)& \multirow{2}{*}{OpenSTT~\cite{OpenSTT}}                       & 3.6 hr    &  3.7k     & 28.1k \\
                                        & test-Y (YouTube)&                             & 3.4 hr                & 3.9k                  & 31.2k \\\hline
        \multirow{4}{*}{English}        & train         & CV-330                        & 330.0 hr              & 208.9k               & 2.2M \\\cline{3-3}
                                        & valid         & \multirow{2}{*}{CV~\cite{DBLP:conf/lrec/ArdilaBDKMHMSTW20}}   & 7.4 hr    & 4.3k      & 43.9k \\
                                        & test          &                               & 7.4 hr                & 4.6k                 & 44.3k \\\cline{3-3}
                                        & test-SF       & SpeakingFaces~\cite{DBLP:journals/corr/abs-2012-02961}        & 7.7 hr    & 6.8k      & 37.7k \\
        \midrule
        \multirow{3}{*}{\textbf{Total}} & train         &  \multirow{3}{*}{-}           & 975.6 hr              & 579.3k                & 6.0M \\
                                        & valid         &                               & 21.6 hr               & 12.4k                 & 127.5k \\
                                        & test          &                               & 29.1 hr               & 22.5k                 & 177.3k \\
        \bottomrule
    \end{tabular}
\end{table}

\subsection{The Kazakh Language}
For Kazakh, we used the recently presented open-source Kazakh Speech Corpus (KSC)~\cite{khassanov-etal-2021-crowdsourced}.
The KSC contains around 332 hours of transcribed audio crowdsourced through the Internet, where volunteers from different regions and age groups were asked to read sentences presented through a web browser.
In total, around 153,000 recordings were accepted from over 1,600 unique devices.
All accepted recordings were manually checked by native Kazakh speakers.
In the KSC, all texts are represented using the Cyrillic alphabet, and audio recordings are stored in the WAV format.
For the training, validation, and test sets, we used the standard split of non-overlapping speakers provided in~\cite{khassanov-etal-2021-crowdsourced}.

\subsection{The Russian Language}
For Russian, we used a manually-cleaned subset extracted from the Russian Open Speech To Text (OpenSTT) dataset~\cite{OpenSTT}.
The OpenSTT is a multidomain (e.g., radio, lectures, phone calls, and so on) dataset consisting of over 20,000 hours of transcribed audio data.
However, the provided transcriptions are unreliable since they were obtained automatically by using ASR systems, YouTube subtitles (user-provided and auto-generated), and so on. 
Clean transcriptions are provided only for the three validation sets from the books, YouTube, and phone calls domains.

To obtain more reliable training data, we hired fluent Russian speakers and manually re-transcribed a randomly selected 334-hour subset of the OpenSTT.
We selected recordings only from the books and YouTube domains.
We named our clean subset OpenSTT-CS334 and its corresponding 334-hour original version OpenSTT-ORG334. 
For the validation set, we randomly selected a 7-hour subset of the OpenSTT-CS334, and left the remaining 327 hours for training.
For the test set, we used the official validation sets of OpenSTT from the books (test-B) and YouTube (test-Y) domains to match the selected training data.

\subsection{The English Language}
For English, we used a 330-hour subset of Mozilla's Common Voice (CV) project~\cite{DBLP:conf/lrec/ArdilaBDKMHMSTW20} that we will further address as the CV-330.
The CV is a multilingual dataset intended for speech technology research and development.
Its construction procedure is similar to the KSC, where volunteers are recruited to read and verify sentences.
The CV-330 consists of validated recordings that received the highest number of up-votes.
For evaluation purposes, we randomly extracted 7-hour subsets from the standard validation and test sets provided in the CV.
Note that the speakers and texts in the training, validation, and test sets are non-overlapping.

We used an additional evaluation set (test-SF) consisting of Kazakh-accented English recordings extracted from the SpeakingFaces dataset~\cite{DBLP:journals/corr/abs-2012-02961}.
SpeakingFaces is a publicly available multimodal dataset comprised of thermal, visual, and audio data streams.
The audios were recorded using a built-in microphone (44.1 kHz) of a web-camera (Logitech C920 Pro HD) at a distance of approximately one meter.
The dataset consists of over 13,000 audio recordings of imperative sentences\footnote{Verbal commands given to virtual assistants and other smart devices such as `turn off the lights', `play the next song', and so on.} spoken by 142 speakers of different races.
We selected recordings spoken by Kazakhs, which resulted in the total of 75 speakers, each uttering around 90 commands.
Since the total size of the selected recordings is insufficient to build robust ASR systems, we use them only for evaluation purposes.
The produced evaluation set is gender balanced (38 females and 37 males), with the average speaker age of 26 years (ranging from 20 to 47).
To the best of our knowledge, this dataset will be the first open-source Kazakh-accented English data, and it will be more suited for assessing the English speech recognition capability of our E2E ASR models.

\section{Speech Recognition Experiments}\label{sec:exp}
In this section, we describe the experimental setup for both the monolingual and multilingual E2E ASR models, as well as the obtained results.
The multilingual and monolingual models were configured and trained similarly for fair comparison.
The results are reported using the word error rate (WER) metric.

\subsection{Experimental Setup}
All E2E ASR systems were trained on the training sets, using the V100 GPUs running on an Nvidia DGX-2 server; hyper-parameters were tuned on the validation sets, and the final systems were evaluated on the test sets (see Table~\ref{tab:dataset}).
For all systems, the input acoustic features were represented as 80-dimensional log Mel filter bank features with pitch computed every 10 ms over a 25 ms window, and the output units were represented using the character-level graphemes.

To train the E2E ASR systems, we used the ESPnet toolkit~\cite{DBLP:conf/interspeech/WatanabeHKHNUSH18} and followed the Wall Street Journal (WSJ) recipe.
The E2E architecture was based on the Transformer network~\cite{DBLP:conf/nips/VaswaniSPUJGKP17} consisting of 12 encoder and 6 decoder blocks. It was jointly trained with the Connectionist Temporal Classification (CTC)~\cite{DBLP:conf/icml/GravesFGS06} objective function under the multi-task learning framework~\cite{DBLP:conf/icassp/KimHW17}.
The interpolation weight for the CTC objective was set to 0.3 and 0.4 during the training and decoding stages, respectively.
For the Transformer module, we set the number of heads in the self-attention layer to 8 each with 512-dimension hidden states, and the feed-forward network dimensions to 2,048.
In addition, a VGG-like convolution module~\cite{DBLP:journals/corr/SimonyanZ14a} was used to pre-process the input audio features before the encoder blocks.
All models were trained for 120 epochs using the Noam optimizer~\cite{DBLP:conf/nips/VaswaniSPUJGKP17} with the initial learning rate of 10 and 25k warm-up steps.
We set the dropout rate and label smoothing to 0.1.
For data augmentation, we used a standard 3-way speed perturbation~\cite{DBLP:conf/interspeech/KoPPK15} with factors of 0.9, 1.0, and 1.1, and the spectral augmentation~\cite{DBLP:conf/interspeech/ParkCZCZCL19}.
We report results on an average model constructed using the last ten checkpoints.

To evaluate the impact of language models (LM) on recognition performance, we built character-level LMs using the transcripts of the training sets.
The LMs were built as a 2-layer long short-term memory (LSTM)~\cite{DBLP:journals/neco/HochreiterS97} network with a memory cell size of 650 each.
We built both monolingual and multilingual LSTM LMs for monolingual and multilingual E2E ASRs, respectively.
The multilingual LSTM LM was trained on the combined training set.
The LSTM LMs were employed during the decoding stage using shallow fusion~\cite{DBLP:journals/corr/GulcehreFXCBLBS15}.
For decoding, we set the beam size to 60 and the LSTM LM interpolation weight to 0.6.
The other hyper-parameter values can be found in our GitHub repository\textsuperscript{\ref{ft:github}}.

\begin{table}[h]
    \caption{The WER (\%) results of monolingual (mono), multilingual (multi), and independent grapheme set (multi-igs) models. The results showing the impact of language model (LM), speed perturbation (SP), and spectral augmentation (SA) are also reported. The average WER is computed by weighting the WERs using the amount of data in the validation and test sets.
}\label{tab:results}
    \renewcommand\arraystretch{1.075}
    \setlength{\tabcolsep}{1.25mm}
    \begin{tabular}{l|cc|ccc|ccc|cc}
        \toprule
        \multirow{2}{*}{\textbf{Model}} & \multicolumn{2}{c|}{\textbf{Kazakh}}  & \multicolumn{3}{c|}{\textbf{Russian}} & \multicolumn{3}{c|}{\textbf{English}}  & \multicolumn{2}{c}{\textbf{Average}}\\\cline{2-11}
                                & valid         & test          & valid         & test-B    & test-Y    & valid         & test  & test-SF       & valid     & test \\
        \midrule
        \textbf{mono}           & 21.5          & 18.8          & 15.2          & 17.2      & 33.7      & 29.8          & 34.6  & 62.0          & 22.0      & 34.3 \\
        +LM                     & 15.9          & 13.9          & 11.5          & 14.5      & 28.8      & 24.7          & 29.1  & 57.7          & 17.3      & 29.7 \\
        +LM+SP                  & 15.3          & 12.7          & 9.8           & 13.4      & 25.5      & 23.1          & 26.7  & 53.9          & 15.9      & 27.3 \\
        +LM+SP+SA               & \textbf{9.4}  & 8.0           & \textbf{7.5}  & \textbf{11.8} & \textbf{21.9} & \textbf{16.3}     & \textbf{18.9}  & 41.6          & \textbf{11.1} & 20.9 \\\hline
        \textbf{multi}          & 20.4          & 16.3          & 13.7          & 16.5      & 31.5      & 28.0          & 32.2  & 56.0          & 20.5      & 31.4 \\
        +LM                     & 15.4          & 12.6          & 11.2          & 14.7      & 28.0      & 23.7          & 27.5  & 51.4          & 16.7      & 27.6 \\
        +LM+SP                  & 14.4          & 11.8          & 10.2          & 13.8      & 25.8      & 22.5          & 26.4  & 48.3          & 15.6      & 26.0 \\
        +LM+SP+SA               & 9.7           & \textbf{7.9}  & 8.2           & 12.5      & 23.3      & 17.1          & 19.9  & 39.5          & 11.7      & 21.1 \\\hline
        \textbf{multi-igs}      & 20.7          & 16.6          & 13.8          & 16.5      & 31.3      & 27.7          & 32.4  & 56.4          & 20.5      & 31.6 \\
        +LM                     & 15.9          & 12.8          & 11.3          & 14.6      & 27.7      & 23.4          & 27.4  & 51.6          & 16.7      & 27.6 \\
        +LM+SP                  & 14.9          & 12.1          & 10.4          & 13.6      & 25.8      & 22.4          & 26.1  & 49.9          & 15.8      & 26.3 \\
        +LM+SP+SA               & 9.7           & \textbf{7.9}  & 8.3           & 12.5      & 23.2      & \textbf{16.3} & \textbf{18.9}  & \textbf{38.3}    & 11.5      & \textbf{20.5} \\
        \bottomrule
    \end{tabular}
\end{table}

\subsection{Experiment Results}
The experiment results for the three languages are given in Table~\ref{tab:results}.
To obtain an average WER over all the languages, we weighted the WERs by the amount of data in the validation and test sets.

\subsubsection{Monolingual Model.}
The results of monolingual models show that applying LMs and data augmentation consistently improves the WER performance for all languages, where an average WER improvement of 13.4\% was achieved on the test sets (from 34.3\% to 20.9\%).
The best WER result for Kazakh is 8.0\% on the test set.
The best WER results for Russian are 11.8\% and 21.9\% on the test-B and test-Y sets, respectively.
Notice that recognizing YouTube recordings (i.e., test-Y) is more challenging than audiobooks (i.e., test-B) since the former contains recordings with the spontaneous speech.
The best WER results for English are 18.9\% and 41.6\% on the test and test-SF sets, respectively.
Presumably, the poor WER performance on the latter is mostly due to the domain mismatch between the training and test-SF sets because the English training set recordings are read by native English speakers, whereas the test-SF recordings are read by native Kazakh speakers.
Moreover, these sets have been collected differently.
The best average WER result on the test sets for the monolingual models is 20.9\%.

We conducted additional experiments to evaluate the quality of our manually-cleaned subset OpenSTT-CS334.
Specifically, we compared the obtained WER results of the monolingual Russian E2E ASR from the previous experiment against a model trained on the original subset OpenSTT-ORG334.
Both models were configured and trained similarly.
The experiment results given in Table~\ref{tab:results_rsc} show that the model trained on our clean subset achieves absolute WER improvement of 6.6\% for the test-B and 3.1\% for the test-Y compared to the model trained on the original subset.
These results demonstrate the utility of our OpenSTT-CS334 subset.

\begin{table}[t]
    \caption{The comparison of WER (\%) results obtained by monolingual Russian ASR models trained on our clean subset OpenSTT-CS334 and corresponding original subset OpenSTT-ORG334.}\label{tab:results_rsc} 
    \renewcommand\arraystretch{1.2}
    \setlength{\tabcolsep}{3.5mm}
    \begin{tabular}{l|c|ccc}
        \toprule
        \multirow{2}{*}{\textbf{Model}} & \multirow{2}{*}{\textbf{Training set}}& \multicolumn{3}{c}{\textbf{Russian}} \\\cline{3-5}
                                        &                                       & \textbf{valid}    & \textbf{test-B}   & \textbf{test-Y} \\
        \midrule
        \multirow{2}{*}{\textbf{mono}\ +LM+SP+SA}   & OpenSTT-CS334             & 7.5               & 11.8              & 21.9 \\
                                                    & OpenSTT-ORG334            & 12.5              & 18.4              & 25.0 \\
        \bottomrule
    \end{tabular}
\end{table}

\subsubsection{Multilingual Model.}
The use of LMs and data augmentation is also effective for the multilingual models, where an average WER improvement of 10.3\% was achieved on the test sets (from 31.4\% to 21.1\%).
The general trend in the WER performances is similar to the monolingual models.
For example, the best WER for Kazakh is 7.9\%, which is very close to the monolingual baseline.
Slightly worse WERs compared to the monolingual baselines are achieved for Russian, with the test-Y set being more challenging than the test-B set.
Likewise, small WER degradations are observed for English, and the performance on the test-SF set is poorer than on the test set.
However, it is important to mention that the multilingual models achieve noticeable WER improvement of 2.1\% on the test-SF set compared to the monolingual baseline (39.5\% versus 41.6\%).
We presume that this improvement is chiefly due to knowledge transfer from the KSC and OpenSTT-CS334 datasets.
The best average WER result on the test sets for the multilingual models is 21.1\%.

Our experiment results show that the multilingual models with the independent grapheme sets (i.e., multi-igs) achieve similar results to the monolingual models for the Kazakh and English languages, and perform slightly worse for Russian.
Notably, it achieves further improvement over the monolingual baseline on the test-SF (38.3\% versus 41.6\%).
Overall, the WER performances of the two grapheme set construction methods are comparable, with the independent grapheme set being slightly better.
The best average WER result on the test sets for the multi-igs models is 20.5\%, which is the lowest of all the E2E ASR models.

\section{Discussion and Future Work}\label{sec:discus}

\textbf{Code-switching.}
There are two major types of code-switching: inter-sentential and intra-sentential.
In the former, the language switch occurs at sentence boundaries, while in the latter, languages switch within sentences, and thus, resulting in a more complex problem.
Our multilingual models can deal only with the inter-sentential cases.
However, in Kazakhstan, intra-sentential code-switching is commonly practiced, especially switching between Kazakh and Russian languages. 
Therefore, our future work will focus on recognizing intra-sentential code-switching utterances by collecting code-switching data or by employing special techniques dedicated to utilizing monolingual data~\cite{DBLP:conf/interspeech/KhassanovXPZCNM19,DBLP:conf/interspeech/ZengKPXC019}. 

\textbf{Kazakh-accented English}.
Since the developed multilingual E2E ASR is intended to be deployed in Kazakhstan, it is important to ensure that it is suited to recognizing Kazakh-accented English utterances.
Our experiment results show that using training datasets uttered by native English speakers leads to suboptimal performance.
Therefore, future work will focus on improving the recognition of Kazakh-accented English utterances, for example, by applying domain adaptation techniques or collecting in-domain training data. 
Although most Kazakhs are fluent in Russian, it would be still interesting to explore the performance of multilingual ASR models on Kazakh-accented Russian utterances.

\textbf{Dataset.}
In our experiments, for each language we employed different datasets with varying acoustic and linguistic characteristics (i.e., collected in different ways, and covering different topics and speaking styles).
As a result, the Russian and English languages turned out to be more challenging than the Kazakh.
Therefore, future work should minimize the impact of domain mismatch between different datasets. This can be achieved by collecting a multilingual dataset under similar conditions or increasing the dataset domain overlap between languages.
In addition, future work should also study the data efficiency--that is, an increase in performance due to the addition of new data, to infer additional data required to achieve further WER improvements.

\section{Conclusion}\label{sec:conclude}
In this paper, we explored multilingual E2E ASR applied to simultaneously recognize three languages used in Kazakhstan: Kazakh, Russian, and English.
Specifically, we developed both monolingual and multilingual E2E ASR models based on the Transformer networks and compared their performances in terms of WER.
To the best of our knowledge, this is the first multilingual E2E ASR work dedicated to these languages.
In addition, we compared the use of two different grapheme set construction methods (i.e., combined and independent).
We also evaluated the impact of language models and data augmentation techniques on the WER performances of the monolingual and multilingual models and found them extremely effective.
Additionally, we introduced two manually-transcribed datasets: OpenSTT-CS334 and test-SF.
The first one is a manually cleaned 334-hour subset extracted from the OpenSTT dataset.
The second one is a 7-hour set of Kazakh-accented English utterances designed to be used for evaluation purposes.
Given that acquiring high-quality speech data is prohibitively expensive, these datasets will be of great use for the speech community both in academia and industry.
Our experiment results show that the multilingual models achieve comparable results to the monolingual models, while having a similar number of parameters.
The best monolingual and multilingual models achieved average WERs of 20.9\% and 20.5\% on the test sets, respectively.
We strongly believe that the conducted experiments and reported findings will benefit researchers planning to build multilingual E2E ASR systems for similar languages, especially from the post-Soviet space.
We also hope our work will encourage future research that leverages the findings and datasets presented in this paper.

\bibliographystyle{splncs04}
%
\bibliography{main}

\end{document}